\documentclass[aps,preprintnumbers,nofootinbib,showpacs]{revtex4}  

\usepackage{amsmath}
\usepackage{graphicx}

\newcommand{\be}{\begin{equation}}
\newcommand{\ee}{\end{equation}}
\newcommand{\ba}{\begin{eqnarray}}
\newcommand{\ea}{\end{eqnarray}}
\newcommand{\nl}{\nonumber \\}

\newcommand{\eq}{{\rm Eq.\,}}
\newcommand{\eqs}{{\rm Eqs.\,}}

\newcommand{\nr}[1]{(\ref{#1})}
\newcommand{\OO}{{\cal O}}
\newcommand{\gsim}{\stackrel{>}{{}_{\sim}}}
\newcommand{\lsim}{\stackrel{<}{{}_{\sim}}}
\newcommand{\mD}{m_{\textrm D}}

\newcommand{\Cs}{C_{\textrm s}}
\newcommand{\CF}{C_\textrm{F}}
\newcommand{\CA}{C_\textrm{A}}
\newcommand{\TF}{T_\textrm{F}}

\renewcommand{\Re}{{\rm Re}\,}

\def\bx{{\bf x}}
\def\by{{\bf y}}
\def\bp{{\bf p}}
\def\bq{{\bf q}}

\def\bpsi{\vec{\psi}}    
\def\CC{\mathcal{C}}
\def\alphas{\alpha_\textrm{s}}
\def\gs{g_\textrm{s}}

\begin{document}

\title{Finite-size effects on the radiative energy loss of a fast parton in hot and dense strongly interacting matter}

\author{Simon Caron-Huot}
\altaffiliation{Institute for Advanced Study, School of Natural Sciences, Einstein Drive, 
Princeton, NJ 08540, USA}
\email{schuot@ias.edu}
\author{Charles Gale}
\affiliation{Department of Physics, McGill University, 3600 rue University, Montr\'eal, QC H3A 2T8, Canada}

\begin{abstract}
We consider finite-size effects on the radiative energy loss of a fast parton moving in a  finite temperature strongly interacting medium, using the light cone path integral formalism put forward by Zakharov. 
We present a convenient reformulation of the problem which makes possible its exact numerical analysis.
This is done by introducing the concept of a radiation rate in the presence of finite-size effects.
This effectively extends the finite-temperature approach
of AMY (Arnold, Moore, and Yaffe) to include interference between vacuum and medium radiation.
We compare results with those obtained in the regime considered by AMY,
with those obtained at leading order in an opacity expansion, and with those obtained deep
in the LPM (Landau-Pomeranchuk-Migdal) regime.
\end{abstract}


\maketitle

\section{Introduction}

The physics of strongly interacting matter in extreme conditions of temperature and density continues to be one of the most active and fascinating fields of contemporary subatomic physics. It straddles nuclear and particle physics, and enjoys an immediate connection with the science of dense stellar objects. The liveliness of this discipline owes in no small part to the vibrant experimental programs and active theoretical  support that characterize it. In this context, it is fair to write that the Relativistic Heavy Ion Collider (RHIC) continues to be a center of activity, and that this facility now boasts several compelling results \cite{white}. For example, the measurement of a strong suppression of hard partons in nucleus-nucleus collisions has been spectacular \cite{RAA},  confirming - and even exceeding - early expectations \cite{gyuwang}. In addition, the empirical success of the hydrodynamical modelling of soft hadronic observables has been an important factor in popularizing the concept of a strongly coupled quark-gluon plasma (sQGP) \cite{sqgp}. 

The experimental measurements of the quenching of hard QCD jets have stimulated the theory community and several approaches based on perturbative QCD (pQCD) have been put forward to explain them. The formalisms differ in the details by which they treat the QCD medium, but all include the coherence built in multiple scatterings: The so-called Landau-Pomeranchuk-Migdal (LPM) effect \cite{LPM}. Recently, some work has been devoted to comparing the different theoretical approaches, to highlight the similarities, and to point out the differences \cite{TECHQM}. Importantly, the medium in a relativistic heavy ion collision is a dynamical one: It rapidly evolves with time. It is therefore important to separate the consequences of the dynamical evolution scenario from those related to basic and fundamental differences in the theory \cite{bass}. In order to highlight as clearly as possible the different assumptions and ingredients going into the different theory approaches, it has been useful to imagine an idealized setting where the basic formalisms can simply be compared without the complications of realistic dynamics: The ``QCD brick''. The brick is a slab of strongly interacting partonic matter, in thermal equilibrium at a temperature $T$. The goal then consists of analyzing the energy loss of a fast parton, propagating in, and interacting with, the medium. With this philosophy in mind, we concentrate in this paper on AMY, a formalism which stems from finite-temperature field theory and treats the QCD medium dynamically. Its basic premises were established in a series of papers written in the last  decade \cite{AMY,JeonMoore}, and some of its phenomenological consequences have also been investigated \cite{bass,Qin:2009bk,AMYpheno}. The original formulation of AMY was in momentum space, and effects owing to the finite size of the emitting region were therefore a challenge to include.  This paper follows the techniques of the light-cone path integral formalism elaborated by Zakharov \cite{ZakharovL,ZakharovBulk}
and which is known to lead to results similar to those of Baier-Dokshitzer-Mueller-Peigne-Schiff (BDMPS) \cite{BDMPS} and incorporate these effects into the AMY description.
The effects of a finite formation time on inelastic radiation rates and on the time-dependent parton momentum distributions are explored, in order  to eventually assess the consequences of this additional consideration on the phenomenology of RHIC and on that of the LHC.

Our paper is organized as follows: In the next section, the basic theoretical building blocks of our approach are laid down.
In what regards radiation rates, we show that in well-defined limits our results coincide with those obtained in other treatments.
Next we explore numerically the effect of the finite formation time on radiation rates, and then on the parton distribution functions.
We explore the dynamics of the radiated gluons, and then summarize and conclude.

\section{Formalism}

\subsection{Starting point}
\label{formalism}

The starting point here will be Eq.~(4) of Ref. \cite{ZakharovBulk}
expressing the total probability for parton $a$,
produced at time $t=0$ with energy $p$,
to emit a pair $b, c$ of bremsstrahlung partons with energies $k$, $p{-}k$:
\be
 \frac{dP^a_{bc}}{dk} = \frac{P^{a(0)}_{bc}(x)}{\pi p}
 \times \Re
\int_0^\infty dt_1 \int_{t_1}^\infty dt_2
\frac{\partial}{\partial \bx} {\cdot}
\frac{\partial}{\partial \by}
 \left[K(t_2,\bx; t_1,\by) - \mbox{(vac)}\right]_{\bx=\by=0}.
 \label{zakharov}
\ee
For the benefit of the reader, this equation will be re-derived 
in the next section.
In \eq (\ref{zakharov}), $K(t_2,\bx; t_1,\by)$ is $[p/(2k(p{-}k))]^2$ times
the propagator associated with the light-cone Hamiltonian
\begin{subequations}
\label{schro}
\ba
  H &=& \delta E(\bp) - i\CC_3,
\\
 \delta E(\bp) &\equiv& \frac{p\bp^2}{2k(p{-}k)}
 + \frac{m_b^2}{2k} + \frac{m_c^2}{2(p{-}k)}  - \frac{m_a^2}{2p},  \label{schroE}
\\
 \CC_3(\bx) &\equiv&
\frac{C_b{+}C_c{-}C_a}{2} v_2(\bx)
+\frac{C_a{+}C_c{-}C_b}{2} v_2(\frac kp \bx)
+\frac{C_a{+}C_b{-}C_c}{2} v_2(\frac{p{-}k}{p}\bx),  \label{factorizedv}
\ea
\end{subequations}
which acts on wave-function $\psi(\bx)$ on the transverse plane;
$m_{a,b,c}$ and $C_{a,b,c}$ are the respective masses (including
thermal corrections) and group theory Casimirs
of the particles involved ($C_A=3$ for gluons and $C_F=\frac43$ for quarks in QCD);
$\CF v_2(\bx)$ is the dipole propagation amplitude per unit
length for a $q\overline{q}$ pair.
$P^{a(0)}_{bc}(x)$ are the DGLAP splitting kernels \cite{DGLAP} for the process
$a\to bc$, with $x=\frac kp$ the longitudinal momentum fraction of particle $b$:
\be
 \label{DGLAP}
 P^{a(0)}_{bc}(x) = \left\{\begin{array}{ll}
 g^2\CF \frac{1+(1{-}x)^2}{x}, & q\to gq \\          
 g^2C_A \frac{1{+}x^4{+}(1{-}x)^4}{x(1{-}x)}, & g\to gg \\
2g^2N_\textrm{F}\TF \left[x^2+(1{-}x)^2\right], & g\to q\overline{q} \\  
e^2 \frac{1+(1{-}x)^2}{x} & e\to \gamma e
\end{array}\right.
\ee
Final states $bc$ and $cb$ are equivalent
and should be counted only once.  In particular, only the region $x<\frac12$ is to
be included in $g\to gg$.
For QCD, $\CF=\frac43$, $\CA=3$ and $N_\textrm{F}\TF=\frac32$.

At this stage, it is appropriate to observe that the formula contains two time integrations,
which, ultimately, correspond to two emission vertices -- one in the matrix element and one in its complex conjugate.
Since no emission occurs before the jet is produced, both
integrations begin at $t=0$.
But since the radiation can be emitted at any time afterwards, they extend to infinity.
Hence \nr{zakharov} contains two non-compact time integrations, which, in general, have
to be performed numerically.

\subsection{Re-organization}

We re-arrange \eq\nr{zakharov} in two steps in order to make it more suitable
for our analysis.  Mainly, we wish to remove the non-compact time integrations.
First, we Fourier-transform it to transverse
momentum space, with $\int_{\bf k}\equiv \int\frac{d^2{\bf k}}{(2\pi)^2}$,
\be
 \frac{dP^a_{bc}}{dk} = \frac{P^{a(0)}_{bc}(x)}{\pi p}
 \times \Re \int_0^\infty dt_1 \int_{t_1}^\infty dt_2
 \int_{\bq,\bp} \bq{\cdot} \bp \left[
  K(t_2,\bq; t_1,\bp) - \mbox{(vac)}\right]. \nonumber
\ee
In Fourier space, $\CC_3$ becomes the Boltzmann-like
collision operator
\ba
  \CC \,\psi(\bp) &=& \int_\bq C(\bq) \left\{
 \frac{C_b{+}C_c{-}C_a}{2}  [\psi(\bp)-\psi(\bp{-}\bq)]\right.
\nl && \left.\hspace{1.7cm}
 +\frac{C_a{+}C_c{-}C_b}{2} [\psi(\bp)-\psi(\bp{+} \frac kp \bq)]\right.
\nl && \left. \hspace{1.7cm}
 +\frac{C_a{+}C_b{-}C_c}{2} [\psi(\bp)-\psi(\bp{+} \frac{p{-}k}p \bq)]\right\}\ ,
 \label{factorized}
\ea
and $C(\bq)$ is related to the dipole amplitude via
$v_2(\bx)\equiv \int_\bq C(\bq) \left(1-e^{i\bq{\cdot}\bx}\right)$.
This function can also be described
as the Casimir-stripped elastic collision rate for a hard particle:
\be
 C(\bq)= \frac{1}{\Cs} (2\pi)^2 \frac{d^2\Gamma_\textrm{el}(\bq)}{d^2\bq}. \nonumber
\ee
No confusion should arise between the function $C(\bq)$ and the Casimirs $C_s$.

The second step is to make the time integrations compact.
At large times, there are no collisions and $K$ depends on time only through phases, which can be integrated exactly.
We use this when integrating by parts:
\ba
 \int_{t_1}^\infty dt_2 K(t_2,\bq; t_1,\bp) &\equiv&
  \int_{t_1}^\infty dt_2\,
  \left[\frac{d}{dt_2} \left(\frac{e^{-i\delta E(\bq) t_2}}
  {-i\delta E(\bq)}\right)\right] e^{i\delta E(\bq)t_2}K(t_2,\bq; t_1,\bp)
 \nl
 &=&
 \frac{1}{-i\delta E(\bq)} K(\infty,\bq; t_1,\bp)
-\frac{1}{-i\delta E(\bq)} K(t_1,\bq; t_1,\bp)
\nl
&& + \int_{t_1}^\infty dt \frac{i}{\delta E(\bq)}
 \CC(t) K(t,\bq; t_1,\bp).  \label{inter1}
\ea
The first, rapidly oscillating, term on the second line disappears, owing to the
conventional convergence factor $e^{-\epsilon t}$ implicit in \eq\nr{zakharov}.
The second term is closely related to the vacuum (DGLAP) radiation, and for the moment
we simply cancel it against the vacuum subtraction.
In fact, this cancellation is not exact and a remainder associated with the
thermal masses will remain;
we postpone its discussion to section \ref{sec:exp}.
For now, only the last term survives, giving
\be
 \frac{dP^a_{bc}}{dk} = \frac{P^{a(0)}_{bc}(x)}{\pi p}
\times \Re
\int_{0}^\infty dt
\int_0^{t} dt_1
\int_{\bq,\bp} 
\frac{i \bq{\cdot}\bp}{\delta E(\bq)} \CC(t)
 K(t,\bq; t_1,\bp).  \nonumber
\ee

In this form, it is natural to
take a derivative with respect to $t$ to define a radiation rate:
\be
 \frac{d\Gamma^a_{bc}(t)}{dk} \equiv \frac{P^{a(0)}_{bc}(x)}{\pi p}
  \times
\Re \int_0^t dt_1 \int_{\bq,\bp}
  \frac{i\bq{\cdot}\bp}{\delta E(\bq)} \CC(t)
 K(t,\bq; t_1,\bp).
 \label{mainrad}
\ee
This is the formula which we shall study in this paper.
It defines a radiation rate in terms of a single, compact, time integration
that may conveniently be computed numerically.
By construction, its content is fully equivalent to that in the BDMPS-Z probability, \eq\nr{zakharov};
its time integral reproduces the latter.
Details of its numerical implementation are given in Appendix \ref{app:rate}.

It is important to mention that a radiation rate is never uniquely defined.
Indeed, the emission time of a given quanta is itself
ambiguous up to a so-called formation time.
Different prescriptions may lead to different time-dependent rates all associated with the same time-integrated probability.
The present prescription is to locate the emission at the time $t$,
which corresponds physically to the time of the last collision that influences the emission.
This is a reasonable choice, as it is located between the two emission vertices.
It is also a particularly natural and convenient choice because it causes the medium-induced
radiation to stop outside the medium,
that is, $\Gamma$ is zero when $\CC$ vanishes.
In particular, for the brick problem, $\Gamma$ is zero after exiting the brick.
Nevertheless, all possible interference effects (including LPM interference), and memory effects,
are included through the non-local kernel $K$.

\section{Physical interpretation of the formula}

We now provide a heuristic derivation of our formula \nr{mainrad}, as promised.
This will help highlight its underlying assumptions and regime of validity.
In this section we also discuss various analytical limits which are related to approximations used in the literature.

\subsection{Heuristic derivation of \eq\nr{zakharov}}
\label{sec:heuristic}

Numerous derivations of formulae similar to
\nr{zakharov} exist, exploiting one formalism or another \cite{BDMPS,ZakharovL,GLV0,AMY,aurzak}.
We believe that the assumptions to be presented here, which did not previously appear with such generality,
suffice for its validity.

The four assumptions we will need to make are:
\begin{itemize}
\item[1] The three ``hard'' particles $a$, $b$ and $c$ propagate eikonally.
\item[2] The opening angle of the radiation is small and its emission vertices are described by
         the leading-order DGLAP expressions.
\item[3] The elastic collisions against the plasma constituents are instantaneous
   compared to the formation time of the radiation.
\item[4] The transverse size of the radiation, at its formation, is small compared
         to medium scales.
\end{itemize}

The meaning of assumption 1 is that the change in the propagation
direction of any of the hard particles,  over a formation time, is small.
This does not necessarily mean that the opening
angle should be small, only that it should stay constant with time.
This will always be true provided $E/T$ is large for all three particles,
and corrections will come in powers of $T/k$, $k$ denoting the smallest of the energies
of the particles.

The meaning of assumption 2 is that the emission itself is mediated by a leading-order
DGLAP vertex.  This does require a small opening angle,
although a parametrically small angle may not be needed.
This also requires that loop corrections be small.
The relevant loops
are presumably controlled by $\alphas$ at the scale $p_\perp$ which enters in  \nr{mainrad}.
An argument for this can be found in Ref. \cite{arnolddogan}, section VI B.
The values of $p_\perp$ that appear in RHIC and LHC contexts will be discussed below.
Incorporating these loop corrections would be difficult at present.

The meaning of assumption 3 is that the coherence time of the force exerted on the hard
particles is short compared to the time scale of the emission process.
This coherence time can be estimated as $\sim 1/\sqrt{kT}$ for the hardest elastic
collisions up to $\sim 1/T$ for plasma-scale ones ($\sim 1/gT$
for a typical soft collision if the plasma is weakly coupled).
On the other hand, formation times scale like
\be
 t\sim \frac{k}{T^2}, \quad \mbox{Bethe-Heitler regime}, \quad
 t\sim \frac{k^{1/2}}{T^{3/2}}, \quad \mbox{LPM regime}.  \nonumber
\ee
Formation times increase with a power of energy, while coherence times do not.
Hence this assumption will always be valid up to $T/k$ corrections.

Assumption 4 means that the medium can be treated as homogeneous in the
\emph{transverse} plane.  Note that no homogeneity assumption is made
in the longitudinal direction.  The case of a general, time-dependent medium
is dealt with in Appendix \ref{app:rate}.

\begin{figure}
\includegraphics[width=11cm]{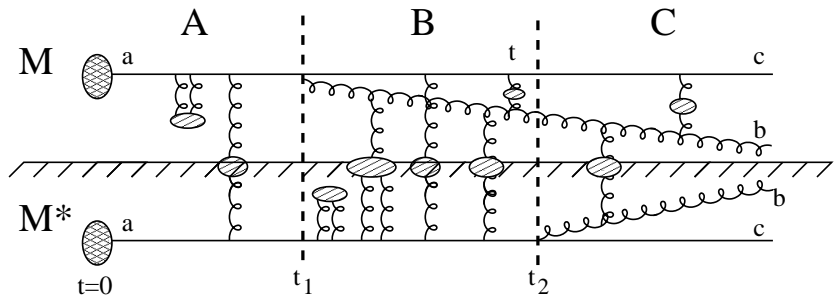}
\caption{A typical matrix element for gluon emission along with its complex conjugate.
  The initial blob represents the hard process that creates the hard parton $a$,
  drawn as a solid line, while the other blobs represent an arbitrary number of plasma particles.
  There is a monotonic time flow from the left to the right of the diagram which is a consequence of the eikonal approximation.
  The collision sub-diagrams, or ``rungs'', are assumed to have negligible horizontal extent.
  Only region B actually contributes to the emission process.
}
\label{fig:mat}
\end{figure}

Within these assumptions, the most general Feynman diagram
contributing to a squared matrix element for bremsstrahlung
is of the ``ladder'' form shown in Figure \ref{fig:mat}.  All such diagrams contribute at the same order
and have to be summed.
The eikonal approximation implies a monotonic time flow along the
horizontal hard lines $a$, $b$ and $c$,
and only two hard vertices connect those lines because of assumption 2.
(One is in the matrix element, the other is in its complex conjugate.)
The vertical rungs describe real and virtual elastic collisions that occur against plasma constituents.
We allow the rungs to be arbitrarily complicated, multi-loop, objects
-- we will only need to assume that they have no horizontal extent.
Because of this, all diagrams are cleanly separated
by the emission vertices into three regions A, B and C.
As we shall now review, only region B contributes to radiation probabilities.

Consider, first, elastic collisions in region C (``elastic'' refers only to the particles in the jet; the plasma particles
 could be scattered very inelastically). 
These collisions will change the momenta of the outgoing
particles and typically reduce their energies by a relatively small amount,
which will modify their momenta measured in the detector.
To discuss the radiation probability, the proper thing to consider is the integrated measurement
over the final momenta.
This is represented symbolically by the operator
\be
  \int_{\bp_b,\bp_c} |\bp_b,\bp_c\rangle \otimes \langle \bp_b,\bp_c| =
 \int d^2\bx_b d^2\bx_c
  ~|\bx_b,\bx_c\rangle\langle \bx_b,\bx_c|
\ee
being convoluted on the right-hand side of the diagram in Figure \ref{fig:mat}.
In the eikonal approximation, this operator is time-independent.
To put it as it is, ``elastic collisions are elastic''.
Hence the sum over states can be performed at time $t_2$, making region C
irrelevant for the radiation probability.
For the same reason, elastic collisions occurring in region A will not affect the emission probability,
as they can be absorbed by a change in the jet axis.

Thus only region B contributes.
In the eikonal approximation, the object that is evolved in that region is a wavefunction of the form
$|\bq_b,\bq_c\rangle \otimes  \langle \bq_a|$,
depending on three transverse momenta.  Equivalently, it depends on three transverse positions.
This describes the eikonal
amplitude times its complex conjugate in region B.
Elastic collisions modify all three momenta but preserve the combination
$(\bq_b{+}\bq_c{-}\bq_a)$.  In fact, this combination vanishes at all times.
Furthermore, it is possible to ``gauge'' to zero
$\bq_a$ (or $\bq_b$ or $\bq_c$)
by a rotation of the beam axis \cite{AMY}.
Provided the opening angles are small, this rotation reduces to linear
shifts $\delta \bq_i \sim p^+_i$, which commute with both the eikonal expression
for the energy $\sim \sum \bq_i^2/p_i$ and with the collision term.
Thus this dynamical gauge fixing does not introduce any new term,
and the wavefunction really depends on a single transverse momentum.
The evolution of this wavefunction will now be expressed in terms of the light-cone Schr\"odinger
equation \nr{schro}.

Choosing to gauge $\bq_a$ to zero so that $\bq\equiv \bq_b=-\bq_c$,
the light-cone energy of the state $|\bq_b,\bq_c\rangle \otimes  \langle \bq_a|$
is
\be 
E_b{+}E_c{-}E_a=
\frac{p\bq^2}{2k(p{-}k) }
 + \frac{m_b^2}{2k} + \frac{m_c^2}{2(p{-}k)}  - \frac{m_a^2}{2p}.  \nonumber
\ee
This gives $\delta E$ in the Schr\"odinger equation \nr{schro}.

The second ingredient in that equation is the collision operator.  Quite generally, within the eikonal
and instantaneous approximations, the vertical rungs in Figure \nr{fig:mat} reduce
to a potential
\be
 -iv(\bx,\frac kp \bx)  \nonumber
\ee
added to the Schr\"odinger equation.
This is imaginary, corresponding to a collision rate.
Here we have used assumption 4, homogeneity in the transverse plane, to reduce
it to a function of two separations, but we could still
be left with a quite complicated function of two arguments.
To proceed further, we need to make the extra assumption
that $v$ is well-described by the dipole-factorized form \nr{factorizedv}:
\be
\frac{C_b{+}C_c{-}C_a}{2} v_2(\bx)
+\frac{C_a{+}C_c{-}C_b}{2} v_2(\frac kp \bx)
+\frac{C_a{+}C_b{-}C_c}{2} v_2(\frac{p{-}k}{p}\bx), \nonumber
\ee
for some function $v_2(\bx)$.
We believe that this is reasonable.  In particular, it was shown in \cite{qhatNLO} that
$\OO(\gs)$ effects at weak coupling do preserve this form.
Perhaps, then, the dominant higher-order effects also do.%
\footnote{
 Incidentally, color structures of the dipole type have been observed to survive
 NLO corrections in the seemingly unrelated context of soft wide-angle radiation \cite{dixonsoft}.
}
In momentum space, this assumption implies that the collision
operator takes the form \nr{factorized} for some function $C(\bq)$.

Since little or no experimental data presently constrains $C(\bq)$,
and since its theoretical determination is so uncertain,
at this stage it is probably best to think of the whole curve
$C(\bq)$ as a phenomenological parameter.
We shall not have much more to say regarding its choice in this paper.

The final ingredients which enter the diagrams
are the hard vertices, which give the DGLAP factor in \nr{zakharov}.

This forms the physical basis of the BDMPS-Z formula.
Diagrammatically, to obtain our equation \nr{mainrad}, one introduces a new time variable
$t$ which corresponds to the time of the last scattering before $t_2$.
The integral over $t_2$ can then be performed analytically, since there are no collisions
between $t$ and $t_2$, and time $t$ becomes the emission time in \nr{mainrad}.

\subsection{Leading order in the opacity expansion}

Gyulassy, Levai and Vitev \cite{GLV0,GLVmain} perform
an expansion in the number  of in-medium collisions, $N$, occurring during a formation time.
Recently, it was shown that this expansion is equivalent
to the series expansion of the BDMPS-Z formula \nr{zakharov} in powers
of $\CC$ \cite{Wicks08}.  Furthermore, its numerical evaluation appears to converge up to $N = 9$.
In that sense the GLV expansion is equivalent to the BDMPS-Z formula.
Our equation \nr{mainrad} provides an exact and numerically tractable implementation of BDMPS-Z,
and its series approximation is thus guaranteed to agree with the GLV series.
For what follows it will be useful to highlight the $N=1$
limit, which will be shown to work very well in certain regimes.

\def\bk{\bf k}

In the $N=1$ approximation, 
$K$ reduces to the free Schr\"odinger propagator,
reducing \nr{mainrad} to
\ba
 \frac{d\Gamma^a_{bc}(t)}{dk} &\to&
 \frac{P^{a(0)}_{bc}(x)p}{4\pi k^2(p{-}k)^2} \int_{\bp}
  \left[\frac{\bp}{\delta E(\bp)}\right]
  \CC 
  \left[\bp \Re i \int_0^t dt_1 e^{-i\delta_E(\bp) (t{-}t_1)} \right]
\nl
&\equiv &
 \frac{P^{a(0)}_{bc}(x)p}{4\pi k^2(p{-}k)^2} \int_{\bq,\bp} C(\bq)
  \frac{1-\cos(\delta E(\bp)t)}{\delta E(\bp)} 
\times \left[
  \frac{C_b{+}C_c{-}C_a}{2}
    \left(\frac{\bp^2}{\delta E(\bp)} - \frac{\bp{\cdot}(\bp{-}\bq)}{\delta E(\bp{-}\bq)}\right)
 + \mbox{2 terms}\right].
\label{GLV1}
\ea
The ``2 terms'' in the square bracket refer to the last two terms in \nr{factorized}.
If we drop the thermal masses in $\delta E$ and make the soft gluon approximation (small $k$),
under which the bracket simplifies considerably,
this precisely reduces to \eq(113) of Ref. \cite{GLVmain}.\footnote{GLV write all results in terms of a scattering length $\lambda$ and
 a $\bq$-distribution of collisions $|\overline{v}(\bq)|^2$ normalized to unity.
 However, only the combination
 $(2\pi)^2|\overline{v}(\bq)|^2/(\lambda C_s)$ appears, which is what we call $C(\bq)$.
 With this substitution (and $Ldz\to dt$), their Eq.(113) literally becomes
 \be
  \frac{dP^a_{ga}}{dk d^2\bk} = \frac{C_a C_g \gs^2}{2\pi^3k}
 \int_0^L dt \int_{\bq} C(\bq)
 \frac{\bk{-}\bq}{(\bk{-}\bq)^2} \cdot\left[ \frac{\bk{-}\bq}{(\bk{-}\bq)^2} - \frac{\bk}{\bk^2}\right]
\left[1-\cos( \frac{(\bk{-}\bq)^2}{2k}t)\right].
  \nonumber
 \ee
 Upon stripping off the $dt$ integration to define a rate and integrating over $\bk$,
 this reproduces \nr{GLV1} with $\bp\to\bk{-}\bq$,
 provided the said approximations are made into the latter.}

In the following,
we will refer to \nr{GLV1} (including thermal masses and without making the soft gluon approximation)
as the $N=1$ approximation.
It is straightforward to evaluate numerically, and it will be seen to be valid at small times.

A word about the collision kernel is in order.  GLV conventionally use the
``static scattering centers'' approximation in which
\be
 C(\bq) = \frac{\gs^4 \sum C_i n_i/d_i}
     {(\bq^2{+}\mu^2)^2}
\ee
with $n_i$, $C_i$ and $d_i$ being the number density, Casimir, and the representation dimension
of various plasma constituents. 
This is a static Yukawa potential, with screening scale $\mu$.
In general, this gets modified by recoil effects, which enhance the elastic cross-sections.
As has been verified recently in great detail \cite{djordjevic} (in the $N=1$ case,
but with a plausible argument for the general case),
a simple change of $C(\bq)$ to \eq\nr{AGZ} given below
will fully account for this effect.
In fact, this particular finding of \cite{djordjevic} is a consequence
of the general argument we have presented above, since the assumptions made in \cite{djordjevic}
are not more general than the ones we have considered.
This result is perfectly in line with the viewpoint advocated in the present paper, which is to view $C(\bq)$
as a phenomenological input to the formalism.
In this work we will only use the collision term \nr{AGZ} below.

\subsection{AMY}

\def\bF{{\bf f}}

Some work has been done, relating the finite-temperature approach of Arnold-Moore-Yaffe (AMY) to that outlined
in section \ref{formalism}, in Ref. \cite{aurzak} and independently in Ref. \cite{arnoldexact}. Other recent work on the AMY formalism includes \cite{SV}. 
The computation in the present subsection is very similar to those in Refs. \cite{aurzak,arnoldexact}.
In both references, the key step was to re-express BDMPS-Z in terms of a rate.
This is also the idea behind \nr{mainrad}.
It is tantamount to formulating AMY in coordinate space.
AMY consider bremsstrahlung from an homogeneous medium of infinite
length and neglect effects associated with the jet production time $t=0$ as well as effects resulting from changes of
the medium during a formation time.
For the brick problem, this means that AMY should be valid in the large $t$ limit.

Our formulation connects very naturally with that of AMY.
In the large $t$ limit \nr{mainrad} becomes
\be
 \frac{d \Gamma^a_{bc}}{dk} \to
 \frac{P^{a(0)}_{bc}(x)p}{4\pi k^2(p{-}k)^2}
\times 
 \Re \int_{\bp} \bF(\bp){\cdot}\bp, \quad
 \label{AMYrad}
  \bF(\bp) \equiv \frac{4k^2(p{-}k)^2}{p^2} \int_{-\infty}^0 dt_1
  \int_\bq \frac{i\bq}{\delta E(\bq)} \circ \CC
    \circ
   K(0,\bq; t_1,\bp).
\ee
Now we observe that
\be
\bF(\bp) + \frac{\bp}{i\delta E} \equiv
 \tilde\bF(\bp) =
\frac{4k^2(p{-}k)^2}{p^2} \int_{-\infty}^0 dt_1
  \int_\bq \bq K(0,\bq; t_1,\bp)
 \nonumber
\ee
has the same real part as $\bF(\bp)$.
By time-translation invariance, $\tilde\bF(\bp)$ obeys the steady-state equation
\be
 \left[i\delta E +\CC\right] \tilde\bF(\bp) = \bp
 \label{schroAMY}
\ee
which suffices to determine it.
\eqs\nr{AMYrad} and \nr{schroAMY} are precisely
equations \nr{AMYrad} and \nr{schroAMY} appearing in \cite{JeonMoore}.%
 \footnote{
  The precise translation is that
  ${\bf h}$ in \cite{JeonMoore} corresponds to
  $p\bp$ here while ${\bf F}$ corresponds to $2p~\tilde\bF$.
}
Hence the rate which we have defined agrees precisely with that defined by AMY when $t\gg t_\textrm{form}$.

It is sometimes stated that the AMY treatment, being rigorously justified
in thermal field theory in the limit of small coupling,
can be valid only for weakly coupled plasmas.
This is incorrect.
We see that
this treatment will always be valid at times larger than a formation time,
within the rather general assumptions discussed in section \ref{sec:heuristic}.
On the other hand, a rigorous application of AMY does suggest the use of 
\be
 C(\bq) = \frac{\gs^2 \mD^2 T}{\bq^2(\bq^2{+}\mD^2)},
\quad\quad
 \mbox{with $\mD^2=\frac32 \gs^2T^2$ in QCD},
\label{AGZ}
\ee
as the collision kernel. 
This expression is obtained at leading-order in the coupling in thermal field theory.
(This was first obtained in this form in \cite{AGZ}.)
This is the only place where a weakly coupled plasma is actually needed, when $k\gg T$
and the conditions of section \ref{sec:heuristic} are satisfied.
Otherwise, if the plasma is not truly weakly coupled but the jet still is,
corrections to \nr{AGZ} are to be expected.

\subsection{The deep LPM regime: The harmonic oscillator}

At asymptotically high energies $p,k\gg T$ the deep LPM regime is realized, in which
a large number of collisions occurs during a formation time.
This motivates a diffusion approximation in which the Hamiltonian becomes
\be
 H = \tilde m_\textrm{eff} + \frac{p\bp^2}{2k(p{-}k)} + \frac{i\hat q_3}{4}
\frac{\partial}{\partial \bp}{\cdot}\frac{\partial}{\partial \bp} \label{Hoscillo},
\ee
where $\tilde m_\textrm{eff}=\frac{m_b^2}{2k} + \frac{m_c^2}{2(p{-}k)}  - \frac{m_a^2}{2p}$.
The effective diffusion coefficient $\hat{q}_3$ is related to the quark momentum broadening
coefficient through $\hat q_3 = \frac{\hat{q}_q}{\CF}\left[
  \frac{C_b{+}C_c{-}C_a}{2}
 +\frac{C_a{+}C_c{-}C_b}{2} \frac{k^2}{p^2}
 +\frac{C_a{+}C_b{-}C_c}{2} \frac{(p{-}k)^2}{p^2}\right]$, where $\hat{q}_q$ is defined as
\be
 \hat{q}_q = \int^{\bq_\textrm{max}} d^2\bq ~\bq^2\frac{d^2\Gamma_\textrm{el}}{d^2\bq}
 = \frac43 \int_\bq^{\bq_\textrm{max}} \bq^2C(\bq).
\label{qhat}
\ee
This Hamiltonian is that of a harmonic oscillator with complex frequency $\omega_0= \sqrt{\frac{-i \hat{q}_3 p}{2k(p{-}k)}}$.
This means that it can be solved analytically.
For a homogeneous brick the exact solution is
\be
 K(t,\bq;t_1,\bp) =
 \frac{p^3}{4k^3(p{-}k)^3} \frac{1}{2\pi i\omega_0 \sin \omega_0 (t{-}t_1)}
\exp \left({i \frac{p[(\bq^2{+}\bp^2) \cos \omega_0 (t{-}t_1)
      -2\bp{\cdot}\bq]}{2k(p{-}k)\omega_0 \sin \omega_0 (t{-}t_1)}
  -i\tilde m_\textrm{eff}(t{-}t_1)
 }\right). \nonumber
\ee
Upon performing the $\bp$ integration in the rate \nr{mainrad} and changing variable to $u\propto \bq^2$,
we obtain
\be
 \frac{d \Gamma^a_{bc}}{dk} \to
 \frac{P^{a(0)}_{bc}(x)}{4\pi^2 k(p{-}k)}
\times \Re \frac{-\hat{q}_3(t) m_\textrm{eff}}{\omega_0} \int_0^t dt'
 \frac{e^{-i\tilde m_\textrm{eff} t'}\sin\omega_0 t'}
 {\cos^3\omega_0 t'}
\int_0^\infty \frac{udu ~e^{-u}} {
  (u + i \frac{m_\textrm{eff}}{\omega_0}\frac{\sin\omega_0 t'}{\cos\omega_0 t'})^3
}.
 \label{oscillo}
\ee
It can be shown that denominators are always non-singular.

The brick result, Eq. \nr{oscillo}, will suffice for the present paper.
Possibly, similar formulae for evolving media could
be derived following the methods of \cite{arnoldexact}.
We will refer to \nr{oscillo} as the ``multiple soft scattering'' approximation.

It is unusual to express results obtained within this approximation
in terms of a rate, and, for this reason, \nr{oscillo} may appear unfamiliar to many readers.
We insist to use a rate in order to facilitate its comparison with \nr{mainrad} later on.
To see that \nr{oscillo} is indeed totally equivalent to the usual results,
consider, for instance, its limit as $m_\textrm{eff}\to 0$:
\be
  \frac{d \Gamma^a_{bc}}{dk} \to
 \frac{P^{a(0)}_{bc}(x)} {8\pi^2 k(p{-}k)} \times \hat{q}_3(t) ~\Re
 \int_0^t dt' \frac{i}{\cos^2(\omega_0 t)}
= \frac{P^{a(0)}_{bc}(x)}{4\pi^2 p} \times \Re \left[-\omega_0 \tan \omega_0 t\right]. \label{masslessosc}
\ee
This is the time derivative of the familiar BDMPS formula
$\frac{d P^a_{bc}}{dk} = \frac{P^{a(0)}_{bc}(x)}{4\pi^2 p} \ln |\cos \omega_0 t|$,
hence it is completely equivalent to it.
More generally, when masses are included, we expect complete agreement with the
formulae of \cite{ZakharovL}.

Physically, this approximation is justified when a large number of soft collisions
occur during a formation time.
While this is formally the case in the high-energy limit,
the corresponding expansion parameter is only $1/\log(E)$ \cite{arnolddogan}.%
 \footnote{The expansion is in logarithms as opposed to powers essentially because
 collisions which are not soft always contribute.
 This is reflected in the logarithmic dependence of the parameter $\hat{q}$ \nr{qhat}
 on an ultraviolet cutoff; collisions near the cutoff are, by definition, not soft.
 Technically, the cutoff behaves like $\bq_\textrm{max} \sim (\hat{q}E)^{1/4}$ \cite{arnolddogan}.}
We will nevertheless verify that this approximation is accurate in the regime
of high energy and sufficiently large time.
At times smaller than a formation time, however, 
this approximation is not well motivated since a large number of collisions may not be reached.
This was emphasized in \cite{zakharov2000,arnold}.
In particular, in \cite{arnold} it is explained that formation
time effects are only well described in a relatively restricted
time window of parametric size $\sqrt{\log(E/T)}$.
This quantity is not large at RHIC (nor will it be at the LHC),
and its smallness will become apparent numerically below.

\section{Multiple emissions}
\label{sec:exp}

So far we have discussed how to calculate the rate at which a single quanta is radiated.
One advantage of a rate formulation is that it is
obvious how to account for multiple emissions: One simply writes down a ``rate equation.''

In this section we keep the notation general, in anticipation of situations in which
the local plasma frame will be evolving with time.
Hence we work entirely in the lab frame.
In general, the evolution of the distribution of quarks and gluons within a jet
may be described within the Boltzmann framework
\ba
  \frac{dn_q(p)}{dt} \!&=&
\!\!\int dk \left[n_q(p{+}k) \frac{d\Gamma^q_{qg}(p{+}k,k)}{dk}
 + 2n_g(p{+}k) \frac{d\Gamma^g_{qq}(p{+}k,k)}{dk}
        - n_q(p)\frac{d\Gamma^q_{qg}(p,k)}{dk}\theta(p{-}k)\right]
\nl &&
  + \textrm{[elastic]}  + \textrm{[vacuum]},
\nl
  \frac{dn_g(p)}{dt}\! &=&
 \!\!\int dk \left[n_q(p{+}k) \frac{d\Gamma^q_{gq}(p{+}k,k)}{dk}
      + n_g(p{+}k) \frac{d\Gamma^g_{gg}(p{+}k,k)}{dk}
      - n_g(p)\left( \frac{d\Gamma^g_{qq}(p,k)}{dk}\theta(p{-}k)
                     +\frac{d\Gamma^g_{gg}(p,k)}{dk}\theta(p{-}2k)\right)
 \right]
\nl &&
  + \textrm{[elastic]}  + \textrm{[vacuum]},
 \label{Boltzmann}
\ea
where the $k$ integrations run from 0 to $\infty$.
The rate $d\Gamma^a_{bc}/dk(p,k)$ for emission of parton $b$ with energy $k$ from
parent parton $a$ with energy $p$ is as defined by \nr{mainrad}.
In an evolving medium its computation is described in the Appendix \ref{app:rate}.
Our notation follows that of \cite{JeonMoore}, although the integration region
is different; \cite{JeonMoore} and other AMY-based treatments typically include absorption processes from the
medium.
The reason we omit these processes will be explained below.

The ``elastic'' component accounts for the energy lost by collisions against plasma constituents.
It has been extensively discussed in the literature \cite{elastic}, and we will not do so here.
We only mention that this component is important when
discussing observables such as $R_{AA}$ \cite{qin}.

It is obvious that a vacuum component has to be included.  This component, however, deserves
further comment.  We must recall what is already included in $d\Gamma^a_{bc}/dk (p,k)$.
This is the vacuum-subtracted rate \nr{mainrad} which includes all interference
effects between vacuum and medium radiation.
What is left is precisely the amount subtracted in \nr{inter1} corresponding to the vacuum radiation,
which may be written
\be
 \frac{dP^{a,\textrm{vac}}_{bc}}{dk} = \frac{P^{a(0)}_{bc}(x)}{2\pi p} \times \int_{\bq}
 \bq^2 \left| \frac{p}{2k(p{-}k)} \int_0^\infty dt e^{-i\int_0^t \delta E} \right|^2. 
 \label{vac}
\ee
In the absence of thermal and vacuum masses, the integral is $\int_{\bq} \frac{1}{\bq^2}$ which
is the standard collinear logarithm for vacuum radiation.
But because thermal masses are nonzero,
\nr{vac} describes the probability for vacuum radiation with a slight suppression
due to the thermal masses.  This suppression is known as the (thermal) Ter-Mikaelian effect
and it has been analyzed in the RHIC context in \cite{adilhorowitz,djordjevicter}.
We will not discuss it further here.

The single-emission probability \nr{vac} is to be exponentiated
so as to maintain a reasonable ordering of events; by our notation we do not wish
to suggest that the vacuum radiation can be naturally handled as a rate.
Some vacuum radiation will occur before or overlap with the earliest medium radiation,
and some will be fragmentation radiation occurring afterwards in the confined phase.
We hope to return to this question of ordering in a future work.
For the application we have in mind in this work, we return to our idealized thermal medium,
and concentrate on the radiative component $d\Gamma^a_{bc}/dk$.

\begin{figure}[t]
\includegraphics[width=8cm]{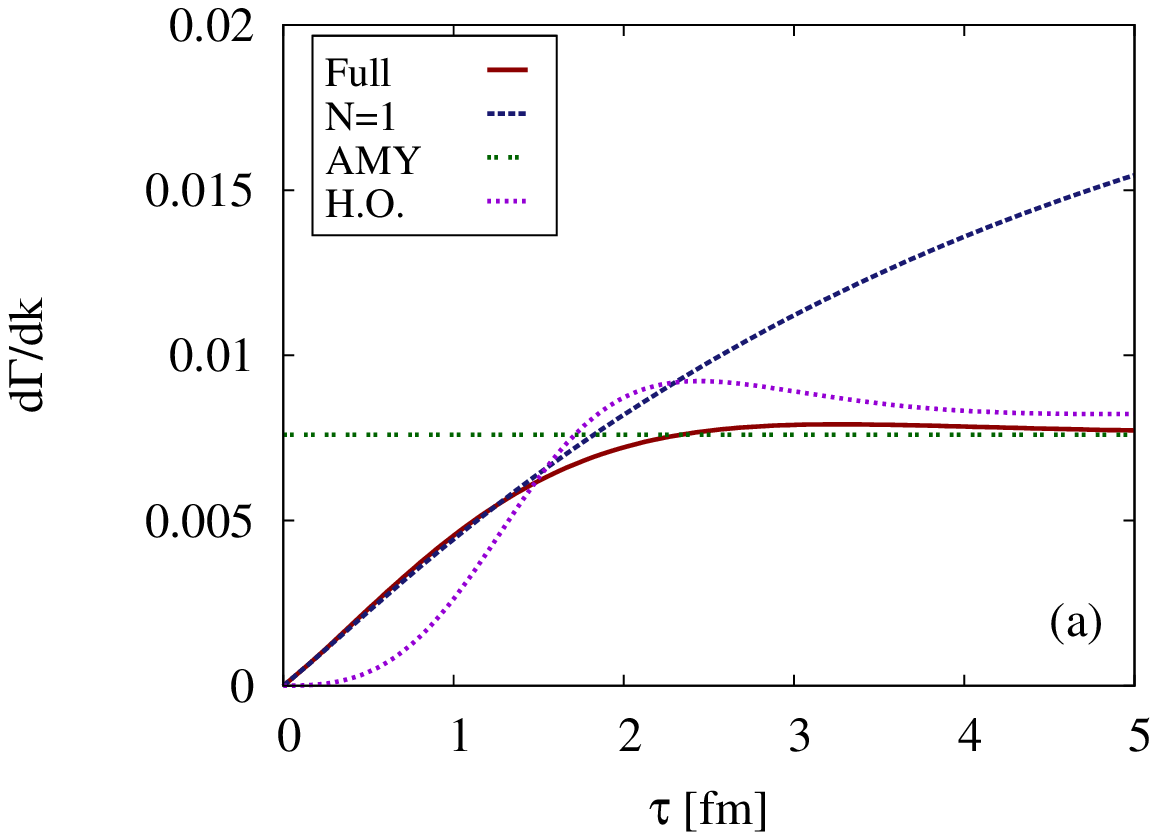}
 \includegraphics[width=8cm]{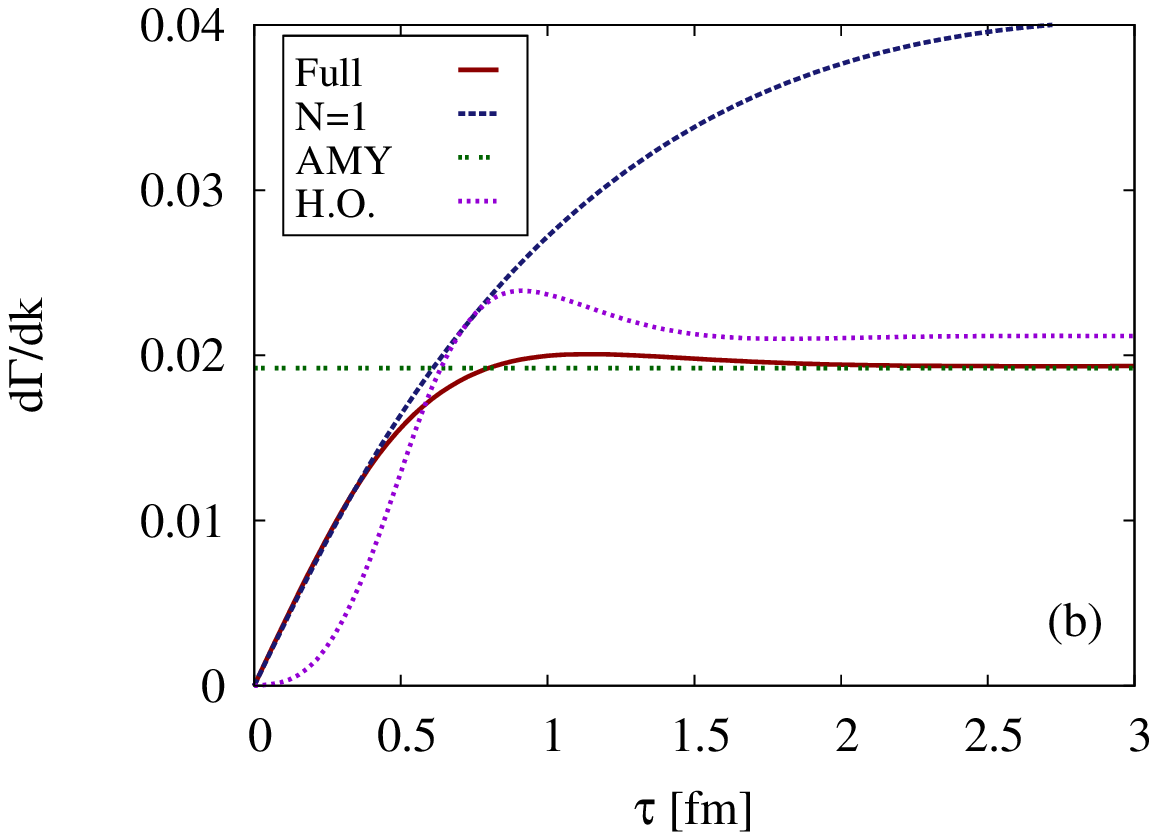}
\caption{(Color online) Radiation rate for a 3 GeV gluon from a 16 GeV parent quark
 as a function of travelled length since the birth of the jet.
  The medium is a uniform brick of QGP at $T=0.2$ GeV (panel (a)) and
  $T=0.4$ GeV (panel (b)), with $\alphas=0.3$.
 }
\label{fig2}
\end{figure}

\begin{figure}[ht]
\includegraphics[width=8cm]{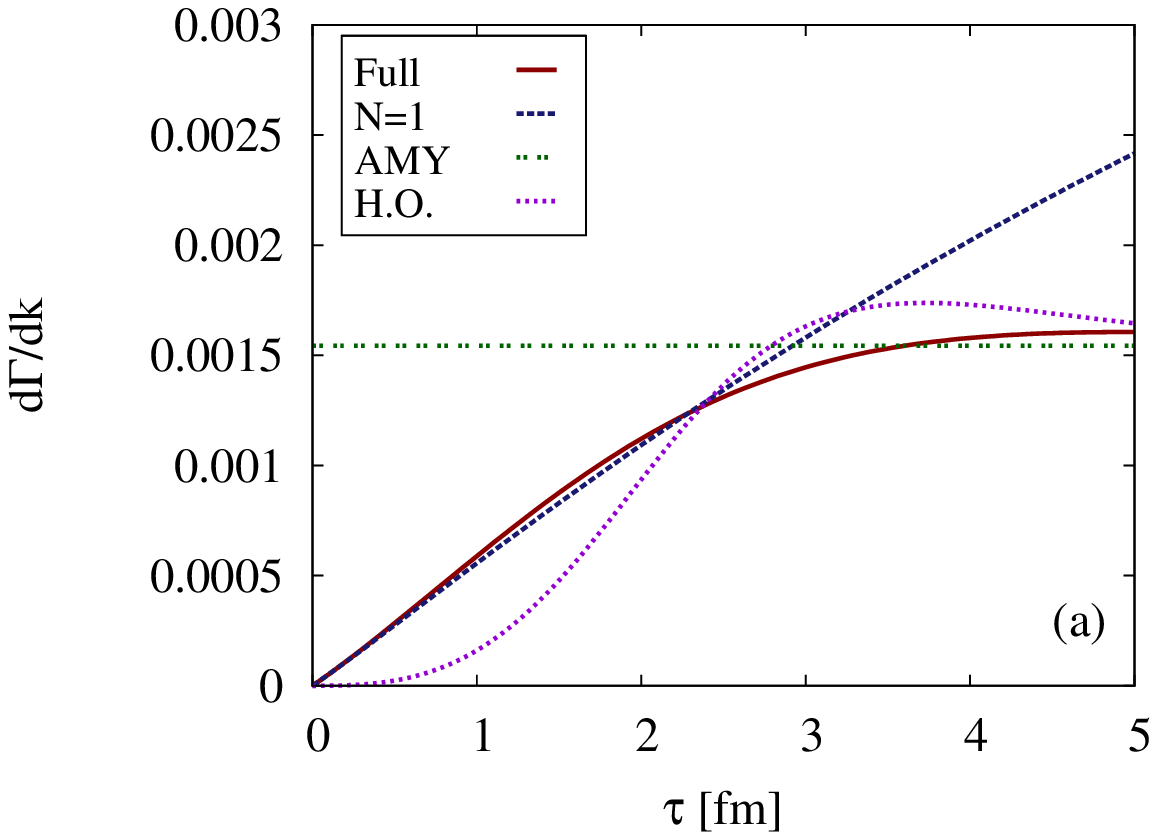}
\includegraphics[width=8cm]{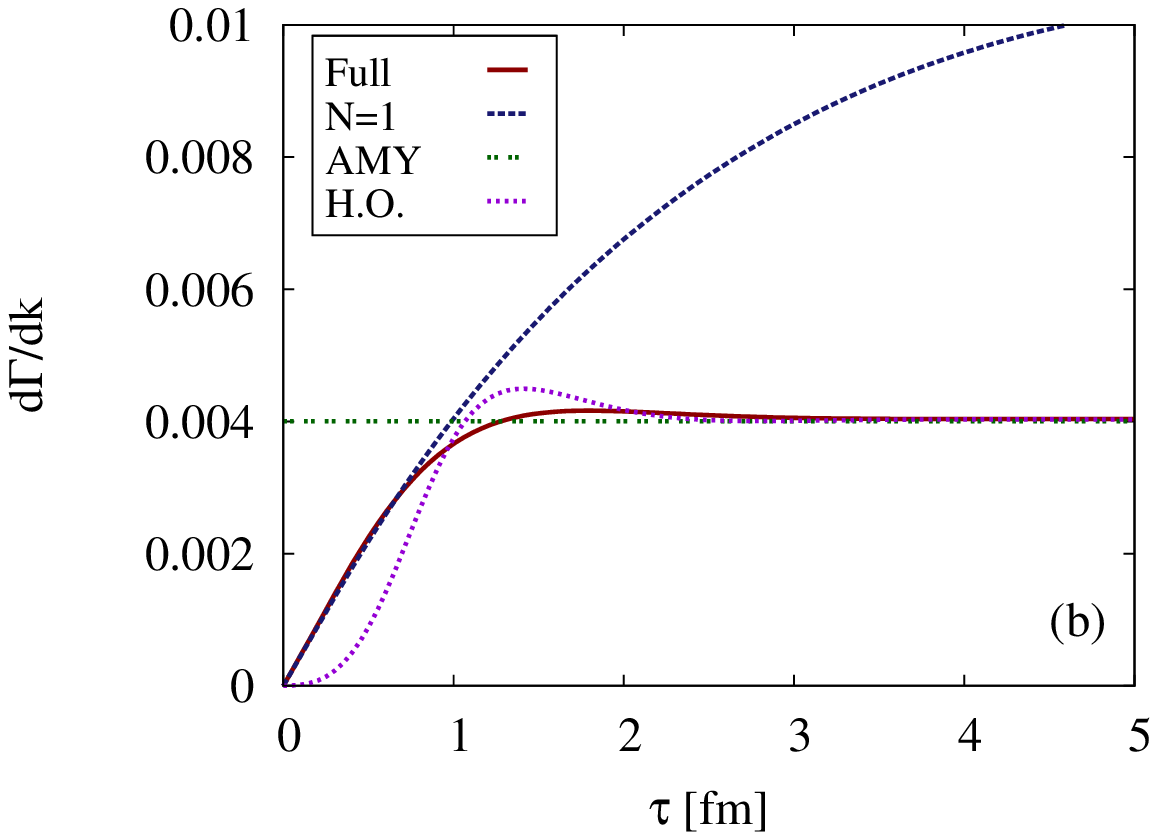}
\caption{(Color online) Same as Figure \ref{fig2} but for a 8 GeV gluon radiated from
  a 16 GeV quark.
 }
\label{fig3}
\end{figure}

\section{Results and discussion}
In this section, we calculate the scattering rates in different approaches
and compare the results. Specifically, we compare results obtained with the
formalism presented here, in which the formation time/length is included explicitly,
with those obtained to first order in the opacity expansion,
with AMY, and in the multiple soft scattering approximation. The first such comparison is shown in Figure \ref{fig2}. 

\begin{figure}[ht]
\includegraphics[width=8.5cm]{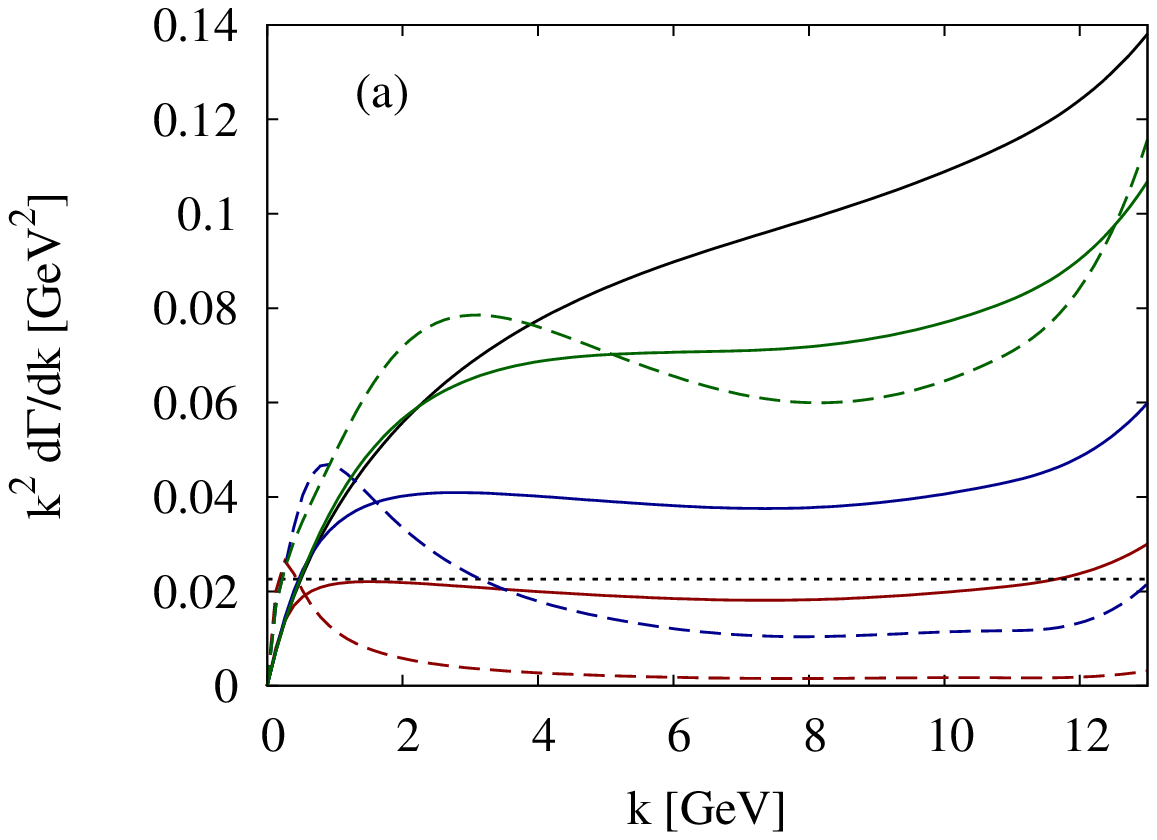}
\includegraphics[width=8.5cm]{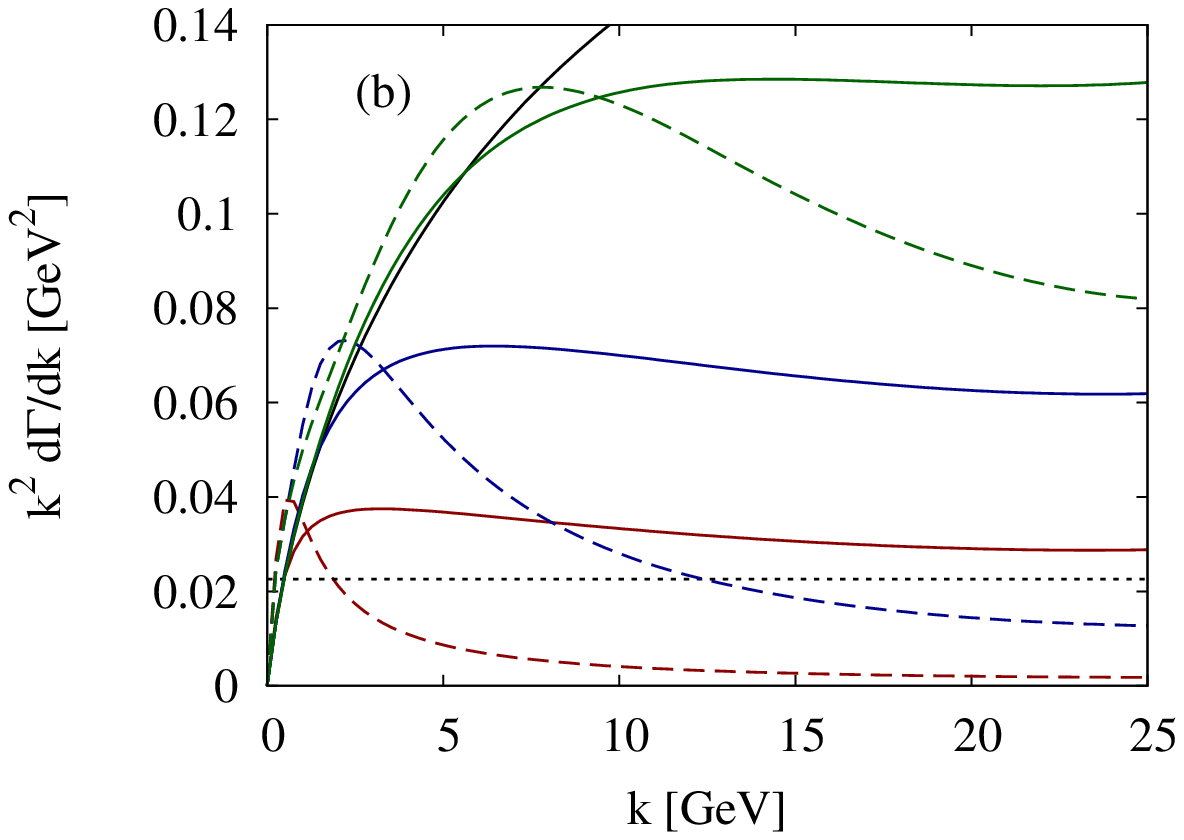}
\caption{(Color online) ``Instantaneous'' radiation spectra $k^2d\Gamma/dk$ in a $T=0.2$ GeV brick for three different times.
    Panel (a): Gluons radiated from a 16 GeV parent quark, at times 0.5 fm/c (red), 1.0 fm/c (blue), and 1.5 fm/c (green). The solid line shows the results of our full calculation, the dashed line the result of multiple soft scatterings, and the dotted line shows the results of purely elastic scatterings. Panel (b): Same as panel (a), but  for a 50 GeV parent quark. Times are 0.8 fm/c (red), 1.6 fm/c (blue), and 3.2 fm/c (green). In both panels, results obtained with AMY are shown as a black solid line.
}
\label{fig:spectra}
\end{figure}

\begin{figure}[ht]
\includegraphics[width=8.5cm]{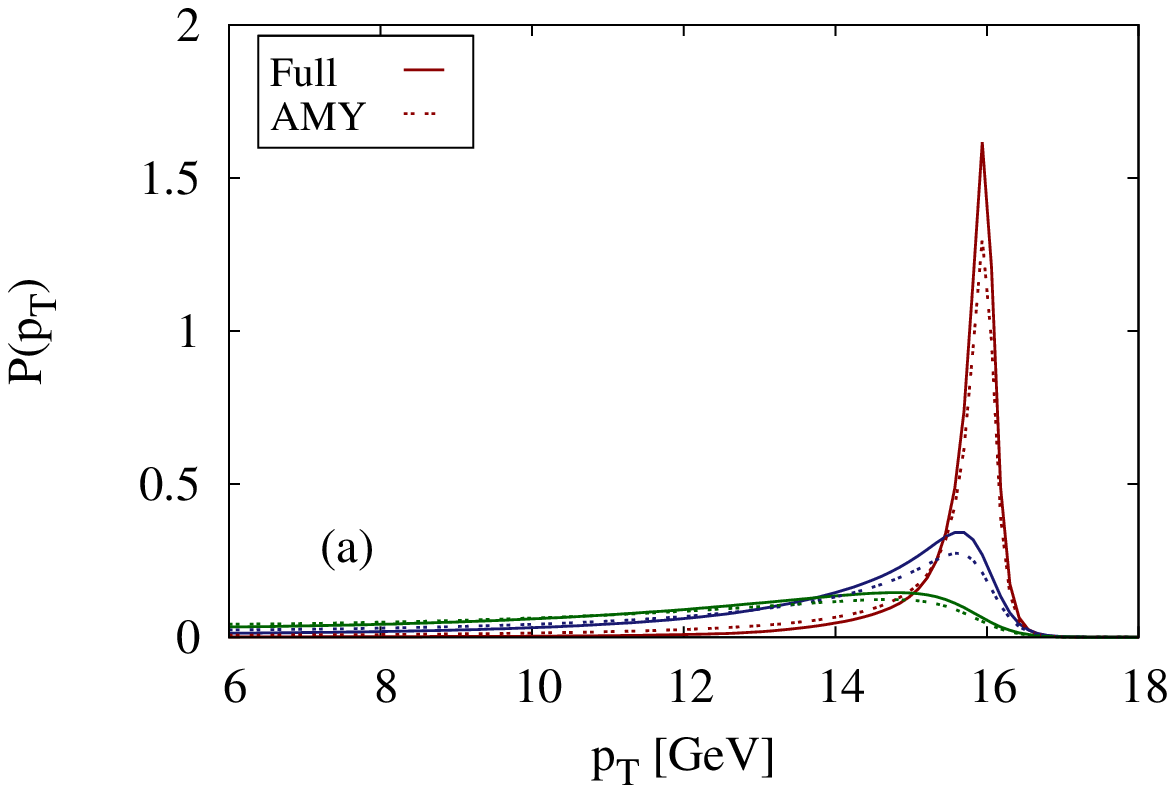}
\includegraphics[width=8.5cm]{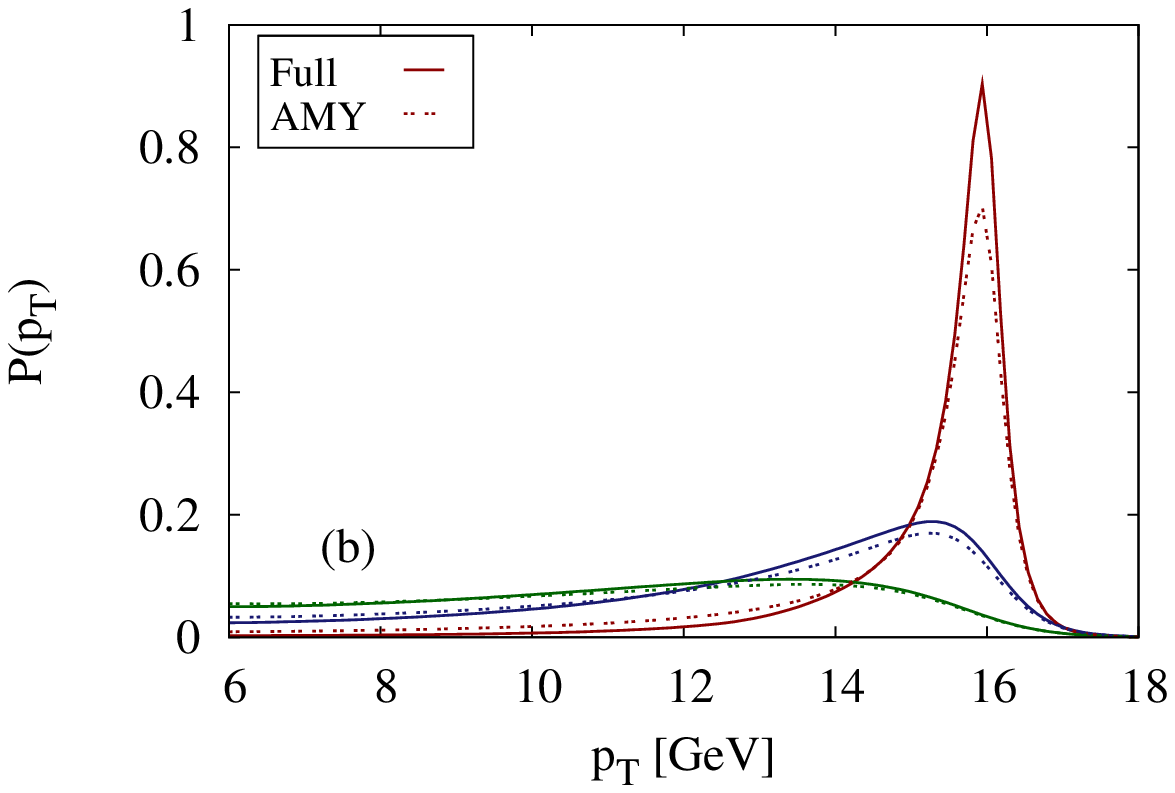}
\caption{(Color online) Degradation of quark transverse momentum as a function of
  time for two QGP bricks at $T=0.2$ GeV (panel (a)),   and $T=0.4$ GeV (panel (b)).  Both initial distributions are
  delta-functions at 16 GeV and are not shown. The three distributions correspond to three different final times.
  From right to left those are: 1, 3, and 5 fm/c for panel (a), and 0.5, 1.5, 2.5 fm/c for panel (b).
}
\label{fig:degrad}
\end{figure}

AMY is seen to be valid for large times,
and our results do tend to that limit as $\tau \to \infty$: A satisfying consistency check.
There is a slight overshoot at a finite time, followed by asymptotic convergence,
which we interpret as a gradual setting in of the LPM suppression.
The left panel of Figure \ref{fig2} shows the differential rate at which a 3 GeV gluon would be radiated off a 16 GeV quark
in a ``brick'' of equilibrated quark-gluon plasma at a fixed temperature of $T = 0.2$ GeV.
The right panel represents results for similar requirements, but for a temperature of 0.4 GeV.
For the coupling constant we used $\alphas=0.3$, which is similar to the value found from experiment in \cite{bass} 
based on the AMY framework\footnote{In fact, preliminary investigations suggest that the phenomenology of one-body observables - like $R_{AA}$ - will require but a modest increase ($\sim$15\%) of $\alpha_s$ \cite{Bjoern_pc}. The relative robustness of single-particle distributions is also apparent from Fig. \ref{fig:degrad}}. 
Figure \ref{fig3} is a repetition of this exercise with different kinematics:
a 16 GeV quark now radiates a 8 GeV gluon.
In both cases, note that the rates obtained here smoothly
interpolate between those obtained at leading order in the opacity expansion 
and those produced by AMY. 

The $N=1$ approximation, \eq\nr{GLV1}, is seen to accurately track the full result at small times.
It begins to deviate at roughly the formation time, due
to the onset of the Landau-Pomeranchuk-Midgal (LPM) interference effect
between multiple scatterings occurring in the medium.
The difference between the $N=1$ and full curves precisely accounts for that effect. 
The amount of that difference is proportional to the 
number of collisions occurring during a formation time and can thus be used to estimate it:
for RHIC energies, we thus estimate that $1 \lesssim N \lesssim 3$, typically.
Higher orders in the opacity expansion have been discussed in \cite{Wicks08}; we have not compared with these results.

The linear rise of rates at small time is thus to be interpreted within the $N=1$ approximation.
At small times, the virtuality of the jet is very high and only very hard collisions
contribute.  In this region, \nr{GLV1} may be estimated as
\be
 d\Gamma/dk \approx P^{(0)}/p \int_{\bq_\textrm{min}}^\infty C(\bq) \propto P^{(0)} g^4n/p\,\bq^2_\textrm{min}
\ee
where $\bq^2_\textrm{min}\sim k/\tau$ and $n$ is a number density of charge carriers.
This produces the linear rise.  This exhibits that the relevant
collisions are harder than the plasma scale, and therefore more weakly coupled,
making this region a robust prediction of QCD.


In Fig \ref{fig:spectra} we indicate the radiation spectrum $k^2d\Gamma/dk$ as a function
of $k$ in a brick of temperature $T=0.2$ GeV, for a $16$ GeV parent quark
with the same parameters as above.  As time increases, the curves move
upwards and eventually saturate the (time-independent) AMY prediction.
In the left panel, this saturation occurs at approximately $t= 3.5$ fm/c.
To avoid making the figures too busy, we have omitted the $N=1$ spectra.
The $N=1$ spectra are indistinguishable from the full curve when the formation time effects
are important, on the right-hand side of the figures (below the AMY curve), but they keep growing
and overshoot the AMY asymptote on the left in keeping with Figures 2 and 3.

\begin{figure}
\includegraphics[width=8cm]{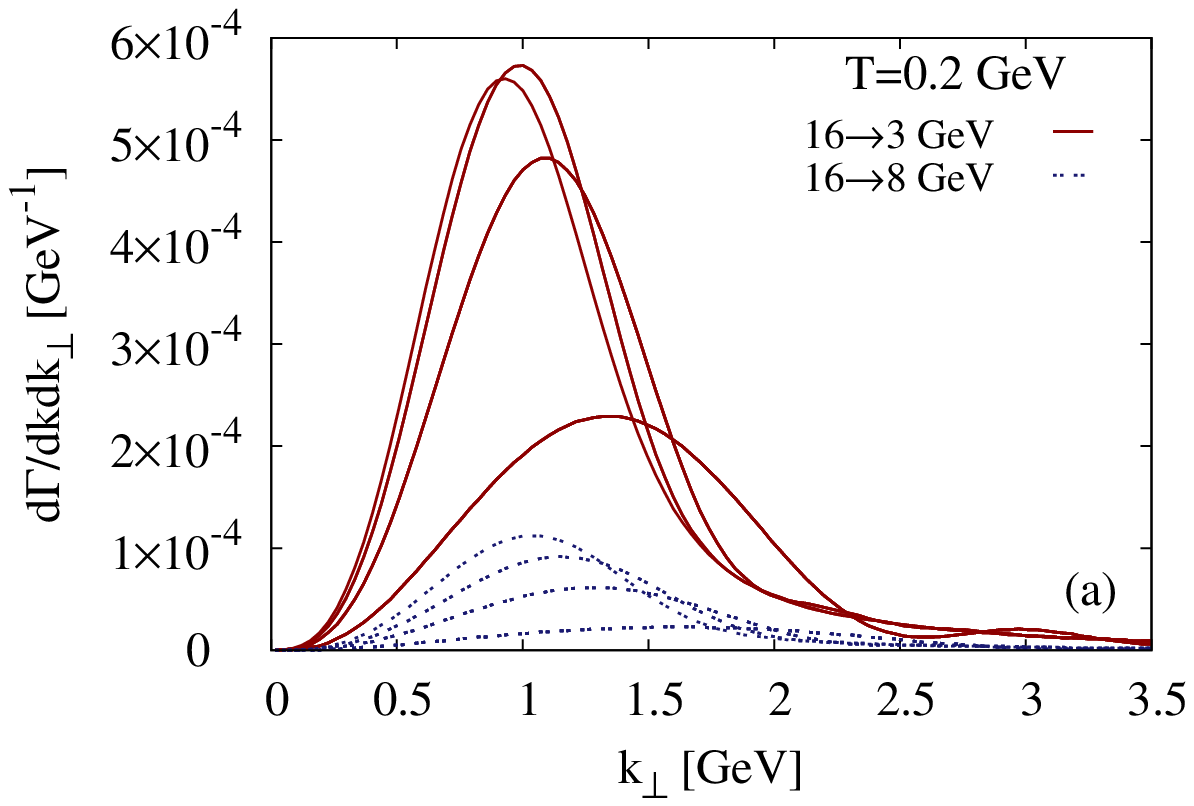}
\includegraphics[width=8cm]{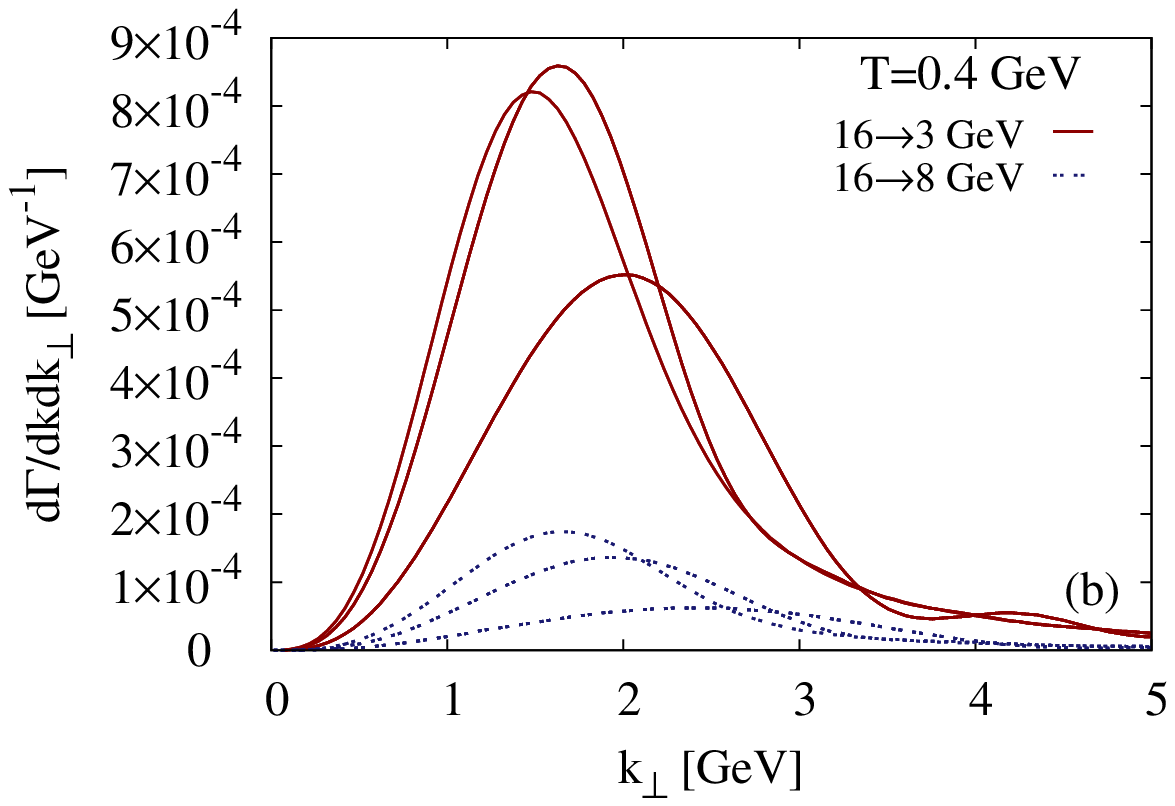}
\caption{(Color online) Transverse momentum distribution of 3 GeV and 8 GeV gluons radiated from a 16 GeV parent quark.
  Panel (a) is for a QGP brick at $T=0.2$ GeV.  From right to left the gluons are emitted
  at times 1, 2, 3 and 5 fm/c.
  Panel (b) is for a QGP brick at $T=0.4$ GeV and times 0.5, 1 and 2 fm/c.
}
\label{fig:qperp}
\end{figure}

Regarding multiple soft approximation, it is important to distinguish between the small and large time regions.
At large times the approximation performs rather well as it tracks closely the AMY result.
It produces slightly red spectra as may be seen in Figure \ref{fig:spectra} but also
in Figure 2, where the oscillator curves asymptote to slightly higher values.
The oscillator curves
depend on a phenomenological parameter $\hat{q}_q$ which we have determined,
for both temperatures, by requiring the correct radiation rate when $k=8$.
That is, the asymptotic agreement in Figure \ref{fig3} is a consequence of our choice of $\hat{q}_q$.
The values we have used are $\hat{q}_q=0.37$ GeV${}^2$/fm
and $2.45$ GeV${}^2$/fm at $T=0.2$ GeV and $T=0.4$ GeV, respectively.
These values should be compared with the definition of $\hat{q}_q$ \nr{qhat}.
For the collision kernel \eq\nr{AGZ}, the latter evaluates to
\be
 \hat{q}_q= \frac{g^2m_D^2T\log(1+\frac{q_\textrm{max}^2}{m_D^2})}{3\pi}.
\ee
Plugging in $q_\textrm{max}\sim 3$ and 4 GeV respectively (estimated from twice
 the values [twice is for $2=p/k$] entering the transverse momentum spectra in Figure \ref{fig:qperp})
 leads to $\hat{q}_q\approx 0.33$ GeV${}^2$/fm and $2.0$ GeV${}^2$/fm respectively.
We conclude that, in the large $t$ region, the values of $\hat{q}$ required by the oscillator model
may be larger than those corresponding to the quark momentum broadening problem but only by a small amount $\lsim 20\%$.

In the small $t$ region, the multiple soft scattering approximation performs poorly
and does not reproduce the full result (which it is supposed to approximate).
This is seen in Figures \ref{fig2} and \ref{fig3} at small times, and is also
is also obvious from Figure \ref{fig:spectra} where the
ultraviolet part of the spectrum is absent at small times.
In general, whenever the ``Full'' and ``AMY'' curves differ, meaning that interference between
vacuum and medium radiation occurs,  the corresponding suppression is exaggerated
by the multiple soft scattering approximation.
Analytically, at fixed $p$ and $k$, the rate in the multiple soft scattering approximation
grows like $t^3$ at small times
while the correct, $N=1$, rate rises like $t$.\footnote{
 The familiar linear rise $dE/dt\propto g^2\hat{q} t$ within the multiple
 soft scattering approximation \cite{BDMPS} is obtained from this
 slow rise $d\Gamma/dk \propto t^3$ at fixed $k$ combined with the rapid motion to the right
 of the peak value $k\propto \hat{q}t^2$.
 As is evident from the figure, in RHIC kinematics the multiple soft scattering contribution
 to $dE/dt$ is swamped out by the $N=1$ contribution which rises linearly as a whole.
}
This qualitative feature has been anticipated in the above-mentioned work \cite{zakharov2000,arnold}.
Finally, we note that in all regions our ``multiple soft approximation''
curves differ by only a very small amount from those obtained within the more familiar
massless approximation \eq\nr{masslessosc}.
All our conclusions thus apply also to the latter, simpler formula.

Our choice to plot the combination $k^2d\Gamma/dk$ is motivated by the fact that this provides
a good estimate for $dE/dt = \int kdk d\Gamma/dk$, which is the relevant concept to discuss the soft
region.  In particular, this quantity is infrared-safe.  At larger values $k\gsim 2\div 3$
GeV (corresponding to the lateral shift of the measured leading-hadron spectra $dN\sim dp_\perp/p_\perp^{8.1}$
     for $R_{AA}\sim 0.2$),
$dE/dt$ ceases to be a useful concept and the more appropriate quantity is $kd\Gamma/dk$.
We have included on the figure the leading-order elastic rate
\be
 \frac{d\Gamma_\textrm{el}}{dk} = \frac{g^2m_D^2C_F}{16 \pi k^2}
\ee
as extracted from \cite{Thoma91}.  Note that there is no kinematic cutoff on the amount of longitudinal energy
which can be lost through elastic scattering.
To the best of our understanding, elastic processes are not interference-suppressed
at small times and so, in the brick problem, produce a time-independent contribution.
As a function of the temperature, we find that the elastic component dominates
the radiative one for $k< 2.5T$.

This may be interpreted in two ways.  First, this means that the elastic component
is only important in the soft region, which is the standard lore.
Second, this implies that soft radiation is largely irrelevant.
Indeed, since there is no invariant way to distinguish
soft emission processes from collisional ones, the best way to look
at radiation with $k\ll 2.5T$ is as a small, sub-leading correction to the elastic component.
This is relevant because \nr{mainrad} was never meant to be accurate in this region anyway.
This is also why we have not included medium absorption processes,
as well as Bose-stimulation or Pauli-blocking factors, in \nr{Boltzmann} --
all these processes account for at most a small increase in the elastic component.

In addition, it is instructive to study the efficiency with which the medium knocks apart the initial delta function, starting for example from a 16 GeV quark.
This is shown in the two panels of Figure \ref{fig:degrad}, for two different brick temperatures. 
At the later times, any hint of a peak structure has mostly been washed out.
Notably, the inclusion of the formation time effects does reduce the distribution in magnitude,
but does not introduce any significant shape distortion.%
  \footnote{In producing the AMY curves in Figure \ref{fig:degrad}, radiation was included only
            for $t>0.7$ fm.  Comparable cutoffs were used in previous AMY implementations \cite{bass},
            where they coincide with the cutoff on the onset of hydrodynamics,
            and the resemblance of this cutoff with the formation times in Figures 2 and 3
            largely explains the apparent smallness of the effect.}

\begin{figure}
\includegraphics[width=8cm]{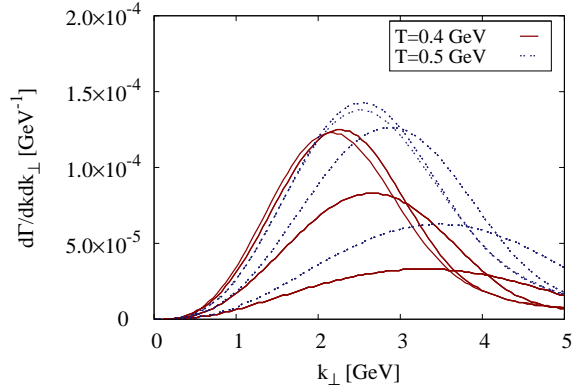}
\caption{(Color online) Transverse momentum distribution of a 10 GeV gluon radiated from a 50 GeV quark at $T=0.4$ GeV and $T= 0.5$ GeV: Conditions of some relevance for the LHC.
  From right to left, times are 0.5, 1, 2, and 3 fm/c.
}
\label{fig:LHC}
\end{figure}

To address the self-consistency of the formalism it will be important to look at
the $k_\perp$ distribution of the radiated gluons.  This is shown in Figure \ref{fig:qperp}.
It is difficult to draw any firm conclusion regarding this issue in the RHIC context
at present since in this paper we have focused on a semi-realistic brick problem.
Nonetheless, in the cases indicated we are satisfied that the relevant values of $k_\perp$ are always
smaller than the energy $k$ of the gluon.  In the worst case we have considered, corresponding
to a 3 GeV gluon radiated from a 16 GeV parent quark in a $T=0.4$ GeV
brick, the typical opening angle from $k_\perp/k$ is approximately $30^\circ$.  In other cases it is much smaller.
The collinear approximation should thus be under control in the cases shown in this paper.
The $k_\perp$ scale is mostly determined by the temperature of the medium,
although $k$ also plays a role (this is in qualitative
 agreement with the parametric estimate $k_\perp\sim (\hat{q} k)^{1/4}$ in the deep LPM regime).
The $k_\perp$ scale in the figure gives the scale at which the
running coupling is to be evaluated
(up to an unknown multiplicative factor whose determination would
 require a full 1-loop computation), which should be
useful for the reliable inclusion of running coupling effects in the future.
In many cases the $k_\perp$'s relevant at RHIC are seen not to be much larger than 1 GeV,
suggesting a possible importance of loop corrections.
On the other hand, strong non-perturbative effects would appear unlikely at such transverse momentum scales.
The situation on all these issues is expected to improve under LHC conditions, as shown in Figure \ref{fig:LHC}.

\section{Conclusion}

We have described a method to solve exactly the BDMPS-Z radiation formula.
The key step in this method is the focus on a differential radiation \emph{rate}
\eq\nr{mainrad}, which is both numerically tractable and physically meaningful.
Such a rate formulation immediately connects with the AMY formalism.
On the other hand, this rate is constructed in such a way that its time integral agrees precisely
with the radiation probability previously defined by BDMPS-Z,
including in situations in which the duration of the emission process is not negligible
and where the AMY treatment was previously not applicable.
This includes interference effects between medium and vacuum radiation.
The method makes it possible
to study numerically and exactly, for the first time, BDMPS-Z radiation spectra in all kinematic regimes.
The rate formulation makes the transitions between different regimes particularly easy to see.

In the brick problem, we have distinguished two different interference-dominated regimes.
In the first one, occurring at small time, the medium modifications to the radiation are determined by
interference between the vacuum radiation and that produced by a single hard collision.
This regime smoothly connects with a second one, in which the vacuum radiation becomes irrelevant
and the radiation is determined by Landau-Pomeranchuk-Migdal (LPM) interference
between in-medium radiation.  The latter regime is well described both by
AMY and by the multiple soft scattering approximation, while the former regime
is well described by the GLV $N=1$ approximation.
The transition between these two regimes occurs
at a physically important length scale, $t\sim 1\div 2$ fm depending upon the energies,
which makes it essential to accurately describe it.

Generally speaking, for a given medium we have shown that
AMY (in previous implementations) over-estimates radiation
while the multiple soft scattering approximation systematically under-estimates it.
This agrees qualitatively with the pattern of interaction rates
which have been extracted from experiment in the past using those models \cite{bass}.

Phenomenologically speaking,
information about the underlying medium enters exactly two places in the present formalism.
One is the elastic collision rate $d^2\Gamma_\textrm{el}(\bq)/d^2\bq$, related to the quantity
$C(\bq)$ in this paper, and the other is the thermal masses.
%
Although we have made no detailed investigation of the effect of thermal masses it seems clear that
the most important parameter is the elastic collision rate.
%
%
For a plasma isotropic in its rest frame, this is a function of $|\bq|$ which is
independent of the jet energy.
Our analysis shows that its second moment, $\hat{q}_q$, does not suffice
to characterize jet energy loss as it ignores a hard Coulomb tail which dominates at small times.
In the present paper we have contented ourselves with the simplest available choice, Eq.~\nr{AGZ},
but it will be important in the future to learn more about the phenomenological implications
of the shape of this curve.

It is known that current efforts to extract meaningful physics from jet quenching
measurements in heavy ion collisions have to be done with the help of hydrodynamic modelling.
The formation time effects
discussed in the present paper are expected to be particularly relevant 
for non-central collisions, and for angular correlation measurements,
where the details of the geometry of the emitting region play a larger role.
As a consequence, a systematic
analysis of several different observables will be needed to quantitatively
characterize the hot and dense strongly interacting medium formed at RHIC,
and soon to be created at the LHC.

\section{acknowledgements}
It is a pleasure to thank N\'estor Armesto, Peter Arnold,  Sangyong Jeon, Abhijit Majumder, Guy D. Moore,
 Carlos Salgado, Bj\"orn Schenke,  and Urs Wiedemann for discussions.
This work was funded in part by the Natural Sciences and Engineering Research Council of Canada, and in part (SCH) by NSF grant PHY-0503584.

\begin{appendix}

\section{Computing the radiation in an expanding medium}
\label{app:rate}

In this appendix, we spell out, for future reference, the proper formulae to be used
in expanding media.   The formula will also be valid in the situations in which the
medium is \emph{not} homogeneous in the longitudinal direction over a formation time.
The derivation follows the discussion in section \ref{sec:heuristic}, where longitudinal homogeneity
was indeed never required.

The generalization of \nr{mainrad} in this case is
\be
 \frac{d\Gamma^a_{bc}(t)}{dk} \equiv \frac{P^{a(0)}_{bc}(x)}{\pi p}
 \times
\Re \int_0^t dt_1 \int_{\bq,\bp}
  \frac{i\bq}{\overline{\delta E}(\bq,t)} \tilde\gamma(t) \,\CC(t)\,
 K(t,\bq; t_1,\bp) \cdot \bp.
 \label{mainradfull}
\ee
In this expression $i/{\overline{\delta E}(\bq,t)}= \int_t^\infty dt' e^{-i\int_t^{t'} \delta E}$.
(This will always be very close to $i/\delta E(\bq,t)$,
 provided thermal masses change slowly.)


Equation \nr{mainrad} contains the time-dilation factor
$\tilde\gamma(t)\equiv \frac{E_\textrm{plasma}}{E_\textrm{lab}}$.
It describes the transformation of $d^2\Gamma_\textrm{el}/d^2\bq$ from the
plasma frame to the lab frame.  This factor always multiply the operator $\CC$.
If, in the lab frame, the plasma is moving with velocity $v$
and at angle $\theta$ relative to the jet axis, this factor may be written
\be
 \tilde\gamma = \frac{1-v\cos\theta}{\sqrt{1-v^2}}. \nonumber
\ee
The expression \nr{factorized} for the now time-dependent $\CC$ is not otherwise modified
and $C(\bq)$ entering it should be computed in the plasma frame as usual.

The same factor multiply $\CC$ in the equation for $K$ which now reads
\be
 \left[\partial_t +i\delta E(\bq,t) + \tilde \gamma(t) \CC(t)\right]
  K(t,\bq;t_1,\bp) =0,
\quad
 K(t_1,\bq;t_1,\bp) =
 \frac{p^2}{4k^2(p{-}k)^2} (2\pi)^2 \delta^2(\bq-\bp).
\ee
The expression for the light-cone energy is unchanged
\be
 \delta E(\bq,t) = \frac{p\bq^2}{2k(p{-}k)}
 + \frac{m_b^2(t)}{2k} + \frac{m_c^2(t)}{2(p{-}k)}  - \frac{m_a^2(t)}{2p}
\ee
provided all energies in it are measured in the lab frame.
For heavy quarks, the
masses are the rest masses, while for light quarks and gluons the masses
are the asymptotic thermal masses (respectively
$m_\infty^2=\frac{\gs^2T^2}{3}$ and
$m_\infty^2=\frac{3\gs^2T^2}{4}$ in QCD at leading order
in thermal perturbation theory).

The solution to these equations exhibits, in an evolving medium,
in addition to the finite-size effects discussed in the present paper,
important memory effects.  Their analysis lies outside our present scope.

We briefly describe our numerical implementation of \nr{mainradfull}.
We view $K(t;t_1)$ as an evolution operator in $t_1$ starting from time $t$.
(Since $t_1<t$, the evolution time is $|t_1{-}t|$.)
The initial wavefunction at time $t$
is the vector-valued function
$
 \bpsi(\bp,|t_1{-}t|{=}0) = \tilde\gamma(t)\CC(t) \circ \frac{i\bp}{\overline{\delta E}(\bp,t)}
$.
We evolve it numerically in the ``interaction picture'' as
\be
 \frac{\partial}{\partial |t_1{-}t|} \bpsi_I(\bp,|t_1{-}t|)
 = -e^{i \int_{t_1}^t \delta E(\bp) } \tilde\gamma(t_1)\CC(t_1)
    e^{-i\int_{t_1}^t \delta E(\bp) } \bpsi_I(\bp,|t_1{-}t|) \nonumber
\ee
Going to the interaction picture removes violent phases
from $\bpsi_I$.
The initial conditions at time $t$
decay at large $\bp$ like $\bpsi_I \sim \bp C(\bp) \sim \bp/\bp^4$.  This
decay is preserved at all times.
To find $\Gamma$ we need to take the integral
$\int_{\bp} \int dt_1 e^{-i\int_{t_1}^t \delta E(\bp) } \,\bp{\cdot}\bpsi_I(\bp,|t_1{-}t|)$.
Because the $\bp$ integral is not absolutely convergent, it is most convenient
to perform the $t_1$ integration first.  At large times, $\psi_I$
tends to zero and the time integral saturates.
However, oscillations in time will cause an extra
suppression $\sim 1/\delta E(\bp)$ at large $\bp$ making the $\bp$ integration convergent,
as it should be expected physically.
Rotational invariance in the $\bp$ plane can be used
to set $\bpsi_I=\frac{\bp}{\bp^2} \tilde{\psi}_I(|\bp|)$ for some scalar
function $\tilde{\psi}_I(|\bp|)$.

\end{appendix}


\begin{thebibliography}{20}
\bibitem{white}M. Gyulassy and L. McLerran, Nucl. Phys. {\bf A750}, 102 (2005); J. Adams {\it et al.,} {\it ibid.}, 184 (2005); I. Arsene {\it et al.,}, {\it ibid.} 1 (2005); B. B. Back {\it et al.,} 28 (2005).
\bibitem{RAA}K. Adcox {\it et al.,} Phys. Rev. Lett. {\bf 88}, 022301 (2002); C. Adler {\it et al.}, Phys. Rev. Lett. {\bf 89}, 202301 (2002).
\bibitem{gyuwang}M. Gyulassy and X.-N. Wang, Nucl. Phys. {\bf B420}, 583 (1994).
\bibitem{sqgp}See, for example,  E.~Shuryak,
  Prog.\ Part.\ Nucl.\ Phys.\  {\bf 62}, 48 (2009), and references therein.
\bibitem{LPM}
L.~D.~Landau and I.~Pomeranchuk, Dokl.\ Akad.\ Nauk Ser.\ Fiz.\ {\bf 92}
 (1953) 535;
L.~D.~Landau and I.~Pomeranchuk, Dokl.\ Akad.\ Nauk Ser.\ Fiz.\ {\bf 92}
 (1953) 735;
A.~B.~Migdal, Dokl.\ Akad.\ Nauk S.S.S.R. {\bf 105}, 77 (1955);
A.~B.~Migdal, Phys.\ Rev.\ {\bf 103}, 1811 (1956).
\bibitem{TECHQM} TEC-HQM Collaboration, in preparation.
\bibitem{bass}S.~A.~Bass, C.~Gale, A.~Majumder, C.~Nonaka, G.~Y.~Qin, T.~Renk and J.~Ruppert,
  Phys.\ Rev.\  C {\bf 79}, 024901 (2009).
\bibitem{AMY}P. Arnold, G. D. Moore, and L. G. Yaffe , JHEP {\bf 0111} 057 (2001); {\bf 0112}, 009 (2001); {\bf 0206}, 030 (2001).
\bibitem{JeonMoore}
  S.~Jeon and G.~D.~Moore,
  Phys.\ Rev.\  C {\bf 71}, 034901 (2005)
  [arXiv:hep-ph/0309332].
\bibitem{Qin:2009bk}
  G.~Y.~Qin, J.~Ruppert, C.~Gale, S.~Jeon and G.~D.~Moore,
  Phys.\ Rev.\  C {\bf 80}, 054909 (2009)
  [arXiv:0906.3280 [hep-ph]].


\bibitem{AMYpheno}S.~Turbide, C.~Gale, E.~Frodermann and U.~Heinz,
  Phys.\ Rev.\  C {\bf 77}, 024909 (2008);   B.~Schenke, C.~Gale and G.~Y.~Qin,
  Phys.\ Rev.\  C {\bf 79}, 054908 (2009);  G.~Y.~Qin, J.~Ruppert, C.~Gale, S.~Jeon and G.~D.~Moore,
  Phys.\ Rev.\  C {\bf 80}, 054909 (2009).

\bibitem{ZakharovL}
  B.~G.~Zakharov, JETP Lett.\ {\bf 63} 952 (1996) [hep-ph/9607440].

\bibitem{ZakharovBulk}
  B.~G.~Zakharov, JETP Lett.\ {\bf 65}, 615 (1997) [hep-ph/9704255].

\bibitem{BDMPS}
  R.~Baier, Y.~L.~Dokshitzer, A.~H.~Mueller, S.~Peigne and D.~Schiff,
  Nucl.\ Phys.\  B {\bf 478}, 577 (1996)
  [arXiv:hep-ph/9604327];  {\bf 483}, 291 (1997)
  [arXiv:hep-ph/9607355].




\bibitem{DGLAP}
V.~N.~Gribov and L.~N.~Lipatov, Sov.\ J.\ Nucl.\ Phys.\ {\bf 15} (1972) 438;
V.~N.~Gribov and L.~N.~Lipatov, Sov.\ J.\ Nucl. Phys. {\bf 15} (1972) 675;
L.~N.~Lipatov, Sov. J. Nucl.\ Phys.\ {\bf 20} (1975) 94;
G.~Altarelli and G.~Parisi, Nucl.\ Phys.\ {\bf B126} (1977) 298;
Yu.~L.~Dokshitzer, Sov.\ Phys.\ JETP {\bf 46} (1977) 641.
\bibitem{GLV0}
  M.~Gyulassy, P.~Levai and I.~Vitev,
  Nucl.\ Phys.\  B {\bf 571}, 197 (2000)
  [arXiv:hep-ph/9907461].
 \bibitem{aurzak}P.~Aurenche and B.~G.~Zakharov,
  JETP Lett.\  {\bf 85}, 149 (2007).

\bibitem{arnolddogan}   
  P.~B.~Arnold and C.~Dogan,
  Phys.\ Rev.\  D {\bf 78}, 065008 (2008)
  [arXiv:0804.3359 [hep-ph]].

\bibitem{qhatNLO}
  S.~Caron-Huot,
  Phys.\ Rev.\  D {\bf 79}, 065039 (2009)  [arXiv:0811.1603 [hep-ph]].
\bibitem{dixonsoft}
  S.~M.~Aybat, L.~J.~Dixon and G.~Sterman,
  Phys.\ Rev.\  D {\bf 74}, 074004 (2006)
  [arXiv:hep-ph/0607309].
\bibitem{GLVmain}
  M.~Gyulassy, P.~Levai and I.~Vitev,
  Nucl.\ Phys.\  B {\bf 594}, 371 (2001)
  [arXiv:nucl-th/0006010].

\bibitem{Wicks08}
  S.~Wicks,
  arXiv:0804.4704 [nucl-th].

\bibitem{djordjevic}
  M.~Djordjevic and U.~Heinz,
   Phys.\ Rev.\  C {\bf 77}, 024905 (2008)
   [arXiv:0705.3439 [nucl-th]];
  Phys.\ Rev.\ Lett.\  {\bf 101}, 022302 (2008)
  [arXiv:0802.1230 [nucl-th]].


  
\bibitem{arnoldexact}P.~Arnold,
  Phys.\ Rev.\  D {\bf 79}, 065025 (2009)
  [arXiv:0808.2767 [hep-ph]]; 

\bibitem{SV}
  S.~V.~Suryanarayana,
  arXiv:1005.5122 [hep-ph].

\bibitem{AGZ}
  P.~Aurenche, F.~Gelis and H.~Zaraket,
  JHEP {\bf 0205}, 043 (2002)
  [arXiv:hep-ph/0204146].
\bibitem{arnold} P.~Arnold,         
  Phys.\ Rev.\ D {\bf 80}, 025004 (2009)
  [arXiv:0903.1081 [nucl-th]].

\bibitem{zakharov2000}   
  B.~G.~Zakharov,
  JETP Lett.\  {\bf 73}, 49 (2001)
  [Pisma Zh.\ Eksp.\ Teor.\ Fiz.\  {\bf 73}, 55 (2001)]
  [arXiv:hep-ph/0012360].


\bibitem{elastic}M. G. Mustafa and M. Thoma, Acta. Phys. Hung. {\bf A22}, 93 (2005); M. G. Mustafa, Phys. Rev. C {\bf 72}, 014905 (2005); S. Wicks and M. Gyulassy, J. Phys. {\bf G34}, S989 (2007), A. Majumder, Phys. Rev. C {\bf 80}, 031902 (2009), X.-N. Wang, Phys. Lett. {\bf B650}, 213 (2007); Bj\"orn Schenke, Charles Gale, and Guang-You Qin, Phys. Rev. C {\bf 79}, 054908 (2009).

\bibitem{qin}G.~Y.~Qin, J.~Ruppert, C.~Gale, S.~Jeon and G.~D.~Moore,
  Phys.\ Rev.\  C {\bf 80}, 054909 (2009)
  [arXiv:0906.3280 [hep-ph].

\bibitem{adilhorowitz}    
  A.~Adil, M.~Gyulassy, W.~A.~Horowitz and S.~Wicks,
  Phys.\ Rev.\  C {\bf 75}, 044906 (2007)
  [arXiv:nucl-th/0606010].

\bibitem{djordjevicter}
  M.~Djordjevic and M.~Gyulassy,
  Phys.\ Rev.\  C {\bf 68}, 034914 (2003)
  [arXiv:nucl-th/0305062].
\bibitem{Bjoern_pc}B. Schenke, private communication.
\bibitem{Thoma91}
  M.~H.~Thoma,
  Phys.\ Lett.\  B {\bf 273}, 128 (1991).







\end{thebibliography}
\end{document}